\documentclass[12pt]{article}     
\usepackage{amsmath,amssymb}
\newcommand{\Mat}[2]{\left(\begin{array}{#1}#2\end{array}\right)}
\newcommand{\mx}[1]{\left(\begin{smallmatrix}#1\end{smallmatrix}\right)}
\newcommand\Cl{\mathfrak{Cl}}

\newcommand{\Hil}{{\mathcal H}}
\newcommand{\ket}[1]{| #1 \rangle}
\newcommand{\bra}[1]{\langle #1 |}

\newcommand{\brkt}[2]{\langle #1 \mid #2 \rangle}
\newcommand\Tr{\mathop\mathrm{Tr}}
\newcommand{\op}[1]{\hat{#1}}
\newcommand\osig{\op{\sigma}}
\newcommand{\Id}{\op{\mathsf 1}}
\newcommand{\Eq}[1]{{\rm Eq.~(\ref{#1})}}
\newcommand{\Sec}[1]{{\rm Sec.~\ref{Sec:#1}}}
\newcommand{\Eqs}[2]{{\rm Eqs.~(\ref{#1},\ref{#2})}}

\newcommand{\x}{\op{\mathfrak x}}
\newcommand{\z}{\op{\mathfrak z}}

\newtheorem{Th}{Theorem}

\newcommand{\Tref}[1]{Th.~\ref{#1}}

\begin{document}
%
\title{Quantum Information Processing with Low-Dimensional Systems}

\author{\em Alexander Yu. Vlasov}

\maketitle
\begin{abstract}
A `register' in quantum information processing --- is composition of $k$ 
quantum systems, `qu$d$its'. The dimensions of Hilbert spaces for one qu$d$it 
and whole quantum register are $d$ and $d^k$ respectively, but we should have 
possibility to prepare arbitrary {\em entangled} state of these $k$ systems.
Preparation and arbitrary transformations of states are possible with 
universal set of quantum gates and for any $d$ may be suggested such 
gates acting only on single systems and neighbouring pairs.  
Here are revisited methods of construction of Hamiltonians for such
universal set of gates and as a concrete new example is considered case 
with qutrits.
Quantum tomography is also revisited briefly.
\end{abstract}



\section{Introduction}
\label{Sec:Intro}

{\em Discrete quantum variables} --- are basic resource in
quantum computing. A {\em qubit} is described by
two-dimensional Hilbert space and systems with higher dimensions 
are also widely used \cite{high}.

Quantum mechanics with {\em continuous variables} may be more understanding
due to a correspondence principle. For example, after change of classical 
momentum $q$ and coordinate $p$  to quantum operators $\op q$, $\op p$ in 
simple Hamiltonians we almost directly may produce correct quantum 
description. 

On the other hand, it is impossible to introduce the $\op p$, $\op q$ operators 
for system with finite-dimensional Hilbert space. Even if for large dimensions
$d \gg 2$ the continuous case could be used as an approximate model of a 
discrete system, it does not seem possible for low dimensions.

In 1928 Weyl suggested a method of quantization, appropriate both for
finite and infinite-dimensional case \cite{WeylGQM}. The basic idea --- 
is to use instead of operators of coordinate $\op q$ and momentum $\op p$ 
they exponents with pure imaginary multipliers and instead of Heisenberg 
commutation relations to write {\em Weyl system} 
\begin{equation}
 \op U = e^{i \alpha \op p},\quad \op V = e^{i \beta \op q},
\quad
 \op U\op V = e^{i \alpha \beta}\op V\op U.
\label{WeylSys}
\end{equation}

An analogue of Weyl-Heisenberg commutation relations \Eq{WeylSys} may be 
written also for discrete quantum variables like qubits. It is recollected 
in \Sec{Weyl} \Eqs{WeylPair}{UVVU}. Due to relevance of considered
scheme for finite-dimensional case with spin-1/2 systems 
(so-called {\em Jordan-Wigner representation}) Weyl wrote:
\begin{quote}
\small
   ``Because of these results I feel certain that the general scheme
of quantum kinematics formulated above is correct. But the
field of discrete groups offers many possibilities which we have
not as yet been able to realize in Nature; ''
\ldots
\end{quote}

Nowadays, due to many applications of the Weyl pair \Eq{WeylPair} 
in quantum computations, error correction, cryptography and tomography the 
note about {\em many possibilities in the field of discrete groups} looks
quite justified.

In the quantum information processing are used many {\em entangled} systems
and in \Sec{nqud} is considered specific constructions with tensor product
of Weyl matrices. 
Due to regular algebraic structure, it is convenient to use such operators
for construction of {\em nonlocal} Hamiltonians for universal sets of quantum 
gates in any dimension. In \Sec{univ} are presented methods of construction
of the sets for any $d \ge 2$ together with an example of Hamiltonians for 
qutrits.

Quantum tomography describes effective measurement procedures for 
quantum systems and ensembles. Weyl pair is also useful tool in 
this area. It is discussed briefly in \Sec{tom}.

\section{Pauli and Weyl matrices}
\label{Sec:Weyl}

The Pauli matrices  $\osig_x = \mx{0&1\\1&0}$,
$\osig_y = \mx{0&-i\\i&0}$, $\osig_z = \mx{1&0\\0&-1}$ with
property
\begin{equation}
 \osig_\nu\osig_\mu + \osig_\nu\osig_\mu= 2 \delta_{\mu\nu},
\label{sigcom}
\end{equation}
may be generalized for $d > 2$ 
using {\em Weyl pair}, {\em i.e.}, two $d \times d$ matrices \cite{WeylGQM}
\begin{equation}
 \op U = \mx{0&1&0&\cdots&0\\0&0&1&\cdots&0\\
 \vdots&\vdots&\vdots&\ddots&\vdots\\0&0&0&\cdots&1\\1&0&0&\cdots&0}\!,
 \quad
 \op V = \mx{1&0&0&\cdots&0\\0&\zeta&0&\cdots&0 \\
 0&0&\zeta^2&\cdots&0 \\ \vdots&\vdots&\vdots&\ddots&\vdots\\
 0&0&0&\cdots&\zeta^{d-1}} 
\label{WeylPair}
\end{equation}
with property
\begin{equation}
\op U\op V = \zeta\op V\op U,
\quad
\zeta^d = 1,
\quad
\zeta = e^{2\pi i /d}.
\label{UVVU}
\end{equation}
In quantum information processing the matrices \Eq{WeylPair} were
widely used at first in theory of quantum error correction codes 
\cite{high,QEC}.   

There are different ways to introduce analogues of {\em three}
Pauli matrices, {\em e.g.},
\begin{eqnarray}
&\op X = \op U,
\quad
\op Y = \zeta^{(d-1)/2} \op U \op V,
\quad
\op Z = \op V.&
\label{btau}\\
&\op X^d = \op Y^d = \op Z^d = \Id ,
\quad
\op X \op Y = \zeta \op Y \op X,
\quad
\op Y \op Z = \zeta \op Z \op Y,
\quad
\underline{\op X \op Z = \zeta \op Z \op X}.&
\label{taucom}
\end{eqnarray}
For $d=2$ with $\zeta = \zeta^{-1} = -1$ we have 
$\op X = \osig_x$, $\op Y = \osig_y$, $\op Z = \osig_z$,
and \Eq{taucom} is reduced to \Eq{sigcom}. 
For $d > 2$ and $\zeta \ne \zeta^{-1}$   
it is necessary to remember about an {\em order} 
({\em e.g.}, $\op Z \op X = \zeta^{-1} \op X \op Z$).

\section{Systems with $n$ qu$d$its}
\label{Sec:nqud}

Hilbert space of the system with $n$ qu$d$its ($d=\dim \Hil_d \ge 2 $)
is tensor product with $n$ terms
$ \Hil^{\otimes n}_d = \underbrace{\Hil_d\otimes\cdots\otimes\Hil_d}_n$.
Let us introduce family with $2n$ operators
\begin{equation}
\begin{split}
\x_{2k-1}  & =  
  {\overbrace{\op Z \otimes\cdots\otimes \op Z}^{k-1}\,}\otimes
 \op X \otimes \overbrace{\Id\otimes\cdots\otimes\Id}^{n-k} \, ,
 \\
 \x_{2k}  & = 
  {\underbrace{\op Z \otimes\cdots\otimes \op Z}_{k-1}\,}\otimes
 \op Y \otimes \underbrace{\Id\otimes\cdots\otimes\Id}_{n-k} \, .
\end{split}
\label{defF}
\end{equation}

For any given dimension $d \ge 2$ operators $\x_k$, $k=1,\ldots,2n$ have properties
\begin{eqnarray}
 & \x_j^d = \Id,
\quad
\x_j \x_k = \zeta \x_k \x_j, \quad j < k,
\quad 
\zeta = e^{2\pi i /d}.
&
\label{ZCom}\\
 &(a_1 \x_1 + a_2 \x_2 + \cdots + a_{2n} \x_{2n})^d =
  a_1^d + a_2^d + \cdots + a_{2n}^d.&
\label{Pow}
\end{eqnarray}
For $d=2$ \Eqs{ZCom}{Pow} define generators of the {\em Clifford algebra} 
$\Cl(2n)$ \cite{ClDir}
\begin{equation}
 \x_j \x_k + \x_k \x_j =  2 \delta_{jk}, \quad
 (a_1 \x_1 + \cdots + a_{2n} \x_{2n})^2 =
  a_1^2 + \cdots + a_{2n}^2.
\label{ClCom}
\end{equation} 
For generalized case $d > 2$ \Eq{ZCom} define  
an {\em algebra of the quantum plane} $A^{2n}_\zeta$ \cite{qgroup}.

\section{Universality}
\label{Sec:univ}
The elements described in Sec.~\ref{Sec:Weyl},\ref{Sec:nqud}
let construct Hamiltonians for universal set of quantum gates
with simple methods of decomposition and useful properties: 
\begin{enumerate}
\item It is set of one- and two-qu$d$it gates (a gate for given
 $\op H$ is $\op G^\tau = e^{-i \op H \tau}$).
\item Two-gates are acting on pairs of {\em neighbouring} systems (qu$d$its).
\item Hamiltonians of the two-qu$d$it gates are {\em diagonal}.
\end{enumerate}
Basic idea \cite{clif,tori,qi02} --- is to start with elements 
$\smash{\x_k\x^\dag_{k+1}}$ and use them for construction of Hamiltonians
of one- and two-qu$d$it gates. In proof of universality are used elements 
generated via commutators \cite{DeuUn,DV95}, but due to \Eq{ZCom} they always  
have property $[\op A,\op B]= (1-\zeta^l)\op A\op B$ with an integer 
$l$ and it produces some simplification. 

It is useful also to exchange 
$\op X \leftrightarrow \op Z^\dag$ in \Eq{defF} and define elements
\begin{equation}
\z_{2k-1}^\dag  = 
  \op X^{\otimes(k-1)}\otimes
 \op Z\otimes \Id^{\otimes(n-k)} ,
 \quad
 \z_{2k}^\dag  = 
  \op X^{\otimes(k-1)}\otimes
 \op Y\otimes \Id^{\otimes(n-k)} 
\label{defZ}
\end{equation}
to make two-qu$d$it operators, like $\op Z_k^\dag \op Z_{k+1}$ 
in \Eq{ZXX} below, diagonal
\begin{equation}
 \z_{2k-1}\z^\dag_{2k} = \op X_k, \quad 
 \z_{2k}\z^\dag_{2k+1} = \op Z_k^\dag \op Z_{k+1}.
\label{ZXX}
\end{equation}
Here is used a brief notation 
$\op X_k \equiv \Id^{\otimes(k-1)}\otimes  \op X\otimes \Id^{\otimes(n-k)}$, {\em etc}.

\smallskip

\noindent{\textbullet\em~Qubits.}
$\op X \equiv \op \sigma_x$, $\op Z \equiv \op\sigma_z$.
The elements \Eq{ZXX} are Hermitian and may be used as Hamiltonians.
The Hamiltonians generate only {\em subgroup} of SU$(2^n)$ and this
subgroup is isomorphic with Spin$(2n)$ \cite{clif}, {\em i.e.}, has only
quadratic dimension.
It is the demonstration of important class of {\em nonuniversal} gates and
has analogues both in optical realizations \cite{KLM} and
in ``fermionic'' implementations \cite{TBferm}.

\begin{Th}
Hamiltonians $\op X_k$ of one-qubit gates together with 
diagonal  Hamiltonians $\op Z_k^\dag \op Z_{k+1}$ of two-qubit gates  are
not universal: they generate only quadratic subgroup of\/ {\rm SU}$(2^n)$
isomorphic to {\rm Spin}$(2n)$.
\end{Th}

It is enough for universality to add two Hamiltonians
of one-qubit gates \cite{clif}
\begin{equation}
 \z_{1} = \op Z_1, \quad \z_1\z_2\z_3 = \op Z_2.
\label{X12}
\end{equation}

\begin{Th}
Hamiltonians $\op X_k$, $\op Z_1$, $\op Z_2$ of one-qubit gates  
together with diagonal Hamiltonians $\op Z_k^\dag \op Z_{k+1}$ of two-qubit gates
 generate universal set of quantum gates.
\end{Th}

\noindent{\em\textbullet~Qu-dits.}
For $d>2$, $\z_k$ and $\z_k\z_{k+1}^\dag$ are not Hermitian, but it is enough
to split each term on two Hermitian parts \cite{tori}.
\begin{Th}\label{unqud}
 Hamiltonians $\op Z_1{+}\op Z^\dag_1$,
$i(\op Z_1{-}\op Z^\dag_1)$, {$\op X_k{+}\op X^\dag_k$}, 
{$i\bigl(\op X_k{-}\op X^\dag_k\bigr)$} of one-qudit gates
together with diagonal Hamiltonians 
$\op Z_k \op Z^\dag_{k+1} {+} \op Z_{k+1} \op Z^\dag_k$, 
\mbox{$i(\op Z_k \op Z^\dag_{k+1}{-}\op Z_{k+1} \op Z^\dag_k)$} 
of two-qudits gates
generate universal set of quantum gates in {\rm SU}$(d^n)$.
\end{Th}

\noindent{\em\textbullet~Qutrits.} 
Let us consider the Hamiltonians
for simplest case of qutrit. Initial (non-Hermitian) matrices here 
\begin{equation}
 \op X = \op U = \mx{0&1&0\\0&0&1\\1&0&0}\!,
\quad
\op Z = \op V = \mx{1&0&0\\0&\omega&0\\0&0&\bar\omega}\!,
\quad \omega = e^{2\pi i /3}.
\label{XZ3}
\end{equation}
Let us construct universal set of quantum gates
using elements from \Tref{unqud} and they linear combinations.
An example of the Hamiltonians for one-qutrit gates:
\begin{equation}
 \mx{1&~0&0\\0&-1&0\\0&~0&0},~
 \mx{1&0&~0\\0&0&~0\\0&0&-1},~
 \mx{0&1&1\\1&0&1\\1&1&0},~
 \mx{~0&~i&-i\\-i&~0&~i\\~i&-i&~0}.
\label{G31}
\end{equation}
It is also possible to use only one two-qutrits Hamiltonian 
\begin{equation}
\op H_d =\ket{0}\bra{0} \otimes \ket{0}\bra{0} +
\ket{1}\bra{1} \otimes \ket{1}\bra{1} +
\ket{2}\bra{2} \otimes \ket{2}\bra{2} 
\label{G32}
\end{equation}
for each neighbouring pair
together with one-qutrit gate $\op X$ instead of two Hamiltonians 
$\op Z_k \op Z^\dag_{k+1} {+} \op Z_{k+1} \op Z^\dag_k$, 
\mbox{$i(\op Z_k \op Z^\dag_{k+1}{-}\op Z_{k+1} \op Z^\dag_k)$}
from \Tref{unqud}.

\section{Quantum tomography}
\label{Sec:tom}

It is useful also to recollect briefly methods of quantum tomography
related with operators introduced above.
Let we have unlimited source of quantum systems with 
{\em unknown} state described by a density matrix $\op\rho$.
A simple set of measurement devices may be described by projectors 
$\op{\mathsf P}_k = \ket{\phi_k}\bra{\phi_k}$. Such device
produces ``click'' with probability
\begin{equation}
 p_k = \Tr(\op{\mathsf P}_k \op\rho) = \bra{\phi_k}\op\rho\ket{\phi_k}. 
\label{pk}
\end{equation}
Which sets of vectors $\ket{\phi_k} \in \Hil_d$ are necessary for complete
reconstruction of any density matrix, if all probabilities $p_k$ \Eq{pk}
are estimated after sufficiently large series of measurements? 
In general, a density matrix may be described by $d^2-1$ real parameters and 
it corresponds to minimal amount of such vectors. 

For $d$ is power of prime number $d=p^m$ exist especial symmetric sets 
based on $d+1$ {\em mutually unbiased bases} (MUB) \cite{pict}. 
Such terminology is used because for any two vectors
in different bases is true $|\brkt{\phi}{\varphi}|^2 = 1/d$.
The construction for power $m>1$ intensively uses theory of {\em Galois fields} 
\cite{pict}, but if $d$ itself is prime, there is quite
visual model \cite{symset}:
\begin{Th}
If dimension $d$ is prime number, the eigenvectors of $d+1$ matrices 
$\op Z, \op X, \op X \op Z, \ldots, \op X \op Z^{d-1}$ produce MUB.
\end{Th}

Eigenvectors of matrix $\op Z$ is simply computational basis $\delta_{kl}$
and $d^2$ eigenvectors of other $d$ matrices $\op X \op Z^n$ have components
$\phi_k = \frac{1}{\sqrt{d}}e^{2 \pi i(a k^2 + b k)/d}$, {\em i.e.},
each such vector is described by two fixed numbers $a,b = 0,\ldots,d-1$.

For qutrit it is four matrices 
$\op Z, \op X, \op X \op Z, \op X \op Z^2$ with 12 eigenvectors
\begin{equation}
\!\!
 \Mat{c|c|c}{1&0&0\\0&1&0\\0&0&1}\!\!,~
\frac{1}{\!\sqrt{3}}\!\Mat{c|c|c}{1&1&1\\
 1&\omega&\bar\omega\\
 1&\bar\omega&\omega}\!\!,~
\frac{1}{\!\sqrt{3}}\!\Mat{c|c|c}{1&1&1\\
 \omega&\bar\omega&1\\
 \omega&1         &\bar\omega}\!\!,~
\frac{1}{\!\sqrt{3}}\!\Mat{c|c|c}{1&1&1\\
 \bar\omega&1&\omega\\
 \bar\omega&\omega&1}\!\!.
\!\!
\end{equation}

For tomography of arbitrary system it is always possible 
to use representation of Hilbert space as tensor product of  
prime dimensions, but at least for power of prime such 
procedure is not optimal \cite{pict,symset}. 

\smallskip

The MUB is yet not maximally symmetric, because scalar
product for elements in {\em different} bases is nonzero, 
but in the {\em same} basis all vectors are orthogonal
and any scalar product is null. 

{\bf SIC-POVM conjecture} \cite{SIC} suggests existence of other symmetric sets: 
{\em in any dimension $d$ exist $d^2$ vectors with property 
$|\brkt{\phi}{\varphi}|^2 = 1/(d+1)$ for any two vectors and all the 
vectors may be produced from a single vector 
$\ket{\phi} \mapsto \op X^a \op Z^b \ket{\phi}$}.
It is interesting, that here again is used $\op X$, $\op Z$ pair.

\section{Conclusion}

In quantum information processing together with qubits may be used 
systems with higher dimension of Hilbert space (qu$d$its) and
continuous quantum variables \cite{cont}. 
Any dimension may have specific properties, say for purpose
of quantum tomography it is useful to distinguish case of
prime dimension ($d=2,3,5,7,11, \ldots$), power of prime
($d=4,8,9, \ldots$), and ``general'' case ($d=6,10,12,\ldots$).

On the other hand, there are general methods discussed above for 
the work with systems in any dimension. Despite of obvious difference 
between quantum system with low dimension and continuous limit, there are 
some tools like Weyl pair, that may provide with useful constructions and
hints in many cases.


\end{document}